\documentclass[conference]{IEEEtran}

\usepackage{amsmath,amssymb,eucal}
\usepackage{xifthen}
\usepackage{mathtools}
\usepackage{enumerate}
\usepackage{microtype}
\usepackage{xspace}
\usepackage{bm}
\usepackage[T1]{fontenc}
\usepackage{fancyhdr}
\usepackage{lastpage}

\newcommand{\mbs}[1]{\bm{#1}}
\newcommand{\vect}[1]{{\lowercase{\mbs{#1}}}}
\newcommand{\mat}[1]{{\uppercase{\mbs{#1}}}}

\newcommand{\T}{{\scriptscriptstyle\mathsf{T}}}
\renewcommand{\H}{{\scriptscriptstyle\mathsf{H}}}

\newcommand{\cond}{\,\vert\,}
\renewcommand{\Re}[1][]{\ifthenelse{\isempty{#1}}{\operatorname{Re}}{\operatorname{Re}\left(#1\right)}}
\renewcommand{\Im}[1][]{\ifthenelse{\isempty{#1}}{\operatorname{Im}}{\operatorname{Im}\left(#1\right)}}

\newcommand{\av}{\vect{a}}

\newcommand{\gv}{\vect{g}}
\newcommand{\hv}{\vect{h}}

\newcommand{\nv}{\vect{n}}

\newcommand{\qv}{\vect{q}}

\newcommand{\sv}{\vect{s}}

\newcommand{\uv}{\vect{u}}
\newcommand{\vv}{\vect{v}}
\newcommand{\wv}{\vect{w}}
\newcommand{\xv}{\vect{x}}
\newcommand{\yv}{\vect{y}}
\newcommand{\zv}{\vect{z}}

\newcommand{\nuv}{\vect{\nu}}
\newcommand{\deltav}{\vect{\delta}}

\newcommand{\Lambdam}{\pmb{\Lambda}}
\newcommand{\Phim}{\pmb{\Phi}}

\newcommand{\Am}{\mat{a}}

\newcommand{\Fm}{\mat{f}}

\newcommand{\Mm}{\mat{M}}

\newcommand{\Qm}{\mat{q}}

\newcommand{\Sm}{\mat{s}}

\newcommand{\Wm}{\mat{w}}

\newcommand{\Bc}{{\mathcal B}}

\newcommand{\Ec}{{\mathcal E}}

\newcommand{\Nc}{{\mathcal N}}
\newcommand{\Oc}{{\mathcal O}}

\newcommand{\Xc}{{\mathcal X}}

\newcommand{\CC}{\mathbb{C}}

\newcommand{\Id}{\mat{\mathrm{I}}}
\newcommand{\one}{\mat{\mathrm{1}}}

\newcommand{\CN}[1][]{\ifthenelse{\isempty{#1}}{\mathcal{N}_{\mathbb{C}}}{\mathcal{N}_{\mathbb{C}}\left(#1\right)}}

\renewcommand{\P}[1][]{\ifthenelse{\isempty{#1}}{\mathbb{P}}{\mathbb{P}\left(#1\right)}}
\newcommand{\E}[1][]{\ifthenelse{\isempty{#1}}{\mathbb{E}}{\mathbb{E}\left(#1\right)}}
\renewcommand{\det}[1][]{\ifthenelse{\isempty{#1}}{\text{det}}{\text{det}\left(#1\right)}}
\newcommand{\trace}[1][]{\ifthenelse{\isempty{#1}}{\text{tr}}{\text{tr}\left(#1\right)}}
\newcommand{\rank}[1][]{\ifthenelse{\isempty{#1}}{\text{rank}}{\text{rank}\left(#1\right)}}
\newcommand{\diag}[1][]{\ifthenelse{\isempty{#1}}{\text{diag}}{\text{diag}\left(#1\right)}}

\DeclarePairedDelimiter\abs{\lvert}{\rvert}
\DeclarePairedDelimiter\norm{\lVert}{\rVert}

\newcommand{\defeq}{\triangleq}

\newtheorem{definition}{Definition}
\newtheorem{theorem}{Theorem}
\newtheorem{lemma}{Lemma}
\newtheorem{assumption}{Assumption}

\newcommand{\varepsilonv}{\bm{\varepsilon}}

\begin{document}

\title{On the Degrees of Freedom of time correlated MISO broadcast channel with delayed CSIT}

\author{\authorblockN{Mari Kobayashi, Sheng Yang}
\authorblockA{
SUPELEC \\
Gif-sur-Yvette, France\\
 {\tt \{mari.kobayashi, sheng.yang\}@supelec.fr}
}
\and
\authorblockN{David Gesbert, Xinping Yi}
\authorblockA{Institut Eurecom\\
Sophia-Antipolis, France\\  
{\tt \{david.gesbert, xinping.yi\}@eurecom.fr}
}
}

\maketitle

\begin{abstract}
  We consider the time correlated MISO broadcast channel where the transmitter 
  has partial knowledge on the 
  current channel state, in addition to delayed channel state information (CSI).
  Rather than exploiting only the current CSI, as the zero-forcing precoding, or
  only the delayed CSI, as the Maddah-Ali-Tse~(MAT) scheme, we
  propose a seamless strategy that takes advantage of both. The achievable
  degrees of freedom of the proposed scheme is characterized in terms of
  the quality of the current channel knowledge. 
\end{abstract}%


\section{Introduction}

In most practical scenarios, perfect channel state information at transmitter (CSIT) may not be available due to the time-varying
nature of wireless channels as well as the limited resource for channel estimation. However, many wireless
applications must guarantee high-data rate and reliable communication in the presence of channel uncertainty. In this paper, we consider such scenario in the context of the two-user MISO broadcast channel, where the transmitter equipped with $m$ antennas wishes to send two private messages to two receivers each with a single antenna. The discrete
time baseband signal model is given by 
\begin{subequations}
\begin{align}
y_t &= \hv_t^\T \xv_t + e_t \\
z_t &= \gv_t^\T \xv_t + b_t, 
\end{align}
\end{subequations}
for any time instant $t$, where $\hv_t,\gv_t \in \CC^{m\times 1}$ are
the channel vectors for user~1 and 2, respectively; $e_t, b_t \sim
\CN[0,1]$ are normalized additive white Gaussian noise~(AWGN) at the
respective receivers; the input signal $\xv_t$ is subject to the power
constraint $\E[ \norm{\xv_t}^2 ] \le P$, $\forall\,t$.  
For the case of perfect CSIT, the optimal multiplexing gain of the channel at hand is two achieved by linear strategies such as zero-forcing (ZF) beamforming. It is also well known that the full multiplexing gain can be maintained under imperfect CSIT if the error in CSIT decreases at the rate $P^{-1}$ as $P$ grows \cite{caire2010multiuser,caire2007required}. Further, in the realistic case where the fading process is correlated with a maximum Doppler frequency shift $0\leq F < \frac{1}{2}$, ZF can achieve a fraction $2(1-2F) $ of the optimal multiplexing gain \cite{caire2010multiuser}. This result somehow reveals the bottleneck of a family of precoding schemes relying only on instantaneous CSIT as the fading speed increases ($F\rightarrow \frac{1}{2}$). 
Recently, a breakthrough has been made in order to overcome precisely such a problem. 
In \cite{maddah2010degrees}, Maddah-Ali and Tse showed a surprising result 
that even completely outdated CSIT can be 
very useful in terms of multiplexing gain. For a system with $m=2$ antennas and two users, the proposed scheme, hereafter called MAT, achieves the multiplexing gain of $\frac{4}{3}$, irrespectively of the fading speed. 
This work shifts the paradigm of broadcast precoding from space-only to space-time alignment. The role of delayed CSIT can then be re-interpreted as a ``feedback'' of the past signal/interference heard by the receiver. This side information enables the transmitter to
perform ``retrospective'' alignment in the space \emph{and} time domain, as demonstrated in different multiuser network systems (see e.g. \cite{FandT_Jafar}). 
Although it exhibits optimal rate scaling behavior, the MAT algorithm is designed based on the worst case scenario where the delayed channel feedback provides no information about the current one. This assumption is over pessimistic as most practical channels exhibit some form of temporal correlation. It should be noticed that MAT does not exploit any current CSIT whereas ZF builds only on the current CSIT.
In fact with a simple selection strategy between ZF and MAT, a multiplexing gain of $\max\{2(1-2F), \frac{4}{3}\}$ is achievable. For either very slowly or very rapidly varying channels, a scheme selection approach is reasonable. Yet, for intermediate ranges of temporal correlation, a question arises as how to best exploit both past channel samples and an estimate of the current one, obtained through a linear prediction.

 In order to model the quality of the current CSIT, we introduce a
 parameter $\alpha$ which indicates the rate of decay of the channel
 estimation error when the transmitted power grows. Thus $\alpha=0$ and
 $\infty$ correspond to no and perfect CSIT respectively. 
We propose a seamless scheme which bridges smoothly
between the two extremal schemes ZF and MAT and we characterize the achievable degrees of freedom.  As it will be shown later, the
proposed scheme combines the ZF and MAT principles into a single multi-slotted protocol which relies on the retransmission and alignement of the {\em residual} 
interference caused by the ZF precoder due to the imperfectness of
current channel state information.

In the following, after a brief presentation of the 
assumptions on the CSI and fading process, we present the proposed
scheme as an extension the MAT principle. The achievable degrees of freedom (DoF) of the
proposed scheme are analyzed afterward. Finally, we interpret the obbtained DoF in a practical temporally correlated fading channel scenario where $\alpha$ can be related to the maximum Doppler shift over the time varying channel.

Throughout the paper, we will use the following notations. 
Matrix transpose, Hermitian
transpose, inverse, and determinant are denoted by $\Am^\T$,
$\Am^{\H}$, $\Am^{-1}$, and $\det[\Am]$, respectively. For any real
number $x$, $[x]$ means $\lfloor x \rfloor + \frac{1}{2}$.

\section{System Model}

For convenience, we provide the following definition on the channel
states.
\begin{definition}[channel states]
  The channel vectors $\hv_t$ and $\gv_t$ are called the states of the
  channel at instant $t$. For simplicity, we also define the state
  matrix $\Sm_t$ as $\Sm_t \defeq \left[ \begin{smallmatrix}\hv_t^\T \\ \gv_t^\T
  \end{smallmatrix} \right]. $
\end{definition}
The assumptions on the fading process and the knowledge of the channel
states are summarized as follows. 
\begin{assumption}[mutually independent fading]\label{assumption:fading}
At any given time instant $t$, the channel vectors for the two users
$\hv_t, \gv_t$ are mutually independent and identically
distributed~(i.i.d.) with zero mean and covariance matrix $\Id_m$. 
Moreover, we assume that $\rank[\Sm_t] = 2$ with probability $1$. 
\end{assumption}

\begin{assumption}[perfect delayed and imperfect current CSI]\label{assumption:CSI}
At each time instant $t$, the transmitter knows the delayed channel states up to
instant $t-1$. In addition, the transmitter can
somehow obtain an estimation $\hat{\Sm}_{t}$ of the current channel state $\Sm_t$, i.e., $\hat{\hv}_t$
and $\hat{\gv}_t$ are available to the transmitter with
\begin{subequations}\label{MMSE}
\begin{align}
  \hv_t &= \hat{\hv}_t + \deltav_{t} \\
  \gv_t &= \hat{\gv}_t + \varepsilonv_{t}
\end{align}
\end{subequations}
where the estimate $\hat{\hv}_t$~(also $\hat{\gv}_t$) and 
estimation error $\deltav_{t}$~(also $\varepsilonv_{t}$) are
uncorrelated and both assumed to be zero mean with covariance
$(1-\sigma^2) \Id_m$ and $\sigma^2 \Id_m$, respectively, with $\sigma^2
\le 1$. The receivers knows perfectly $\Sm_t$ and $\hat{\Sm}_t$ without delay.  
\end{assumption}
 
Without loss of generality, we can introduce a parameter $\alpha_P\ge0$
as the power exponent of the estimation error
\begin{align}
  \alpha_P \defeq -\frac{\log(\sigma^2)}{\log P}. \label{eq:alpha}
\end{align}
The parameter $\alpha$ can be regarded as the quality of the
current CSI in the high SNR regime. Note that $\alpha_P=0$ corresponds to the case with no current CSIT at all while $\alpha_P\to\infty$ corresponds to the case with perfect current
CSIT.

\section{Proposed Scheme}

In this section, we propose a novel scheme that combines ZF exploiting
some estimated current CSIT and MAT exploiting delayed CSIT. We start by
briefly reviewing the MAT scheme. 

\subsection{MAT Alignment Revisited}

In the two-user MISO case, the original MAT is a three-slot
scheme, described by the following equations
\begin{subequations}
\begin{align}
  \xv_1 &= \uv & 
  \xv_2 &= \vv & 
  \xv_3 &= [\gv_1^\T \uv + \hv_2^\T \vv  \quad 0]^\T
  \label{eq:tmp201}\\
  y_1 &= \hv_1^\T \uv & 
  y_2 &= \hv_2^\T \vv & 
  y_3 &= h_{31}(\gv_1^\T \uv + \hv_2^\T \vv)  \\
  z_1 &= \gv_1^\T \uv &
  z_2 &= \gv_2^\T \vv &
  z_3 &= g_{31}(\gv_1^\T \uv + \hv_2^\T \vv) 
  \label{eq:tmp203} 
\end{align}
\end{subequations}
where $\uv,\vv\in\mathbb{C}^{m\times 1}$ are useful signals to user~1
and user~2, respectively; for simplicity, we omit the noise in the
received signals.  
The idea of the MAT scheme is to use the delayed CSIT to align the mutual
interference into a reduced subspace with only one dimension~($\hv_1^\T
\vv$ for user 1 and $\gv_1^\T \uv$ for user 2). And importantly, the
reduction in interference is done without sacrificing the dimension of
the useful signals. 
Specifically, a two-dimensional interference-free observation
of $\uv$~(resp. $\vv$) is obtained at receiver~1~(resp.~receiver~2).

Interestingly, the alignment can be done in a different manner.
\begin{subequations}
\begin{flalign}
   \xv_1 &= \uv + \vv& 
   \xv_2 &= [\hv_1^\T \vv \quad 0 ]^\T & 
   \xv_3 &= [\gv_1^\T \uv \quad 0 ]^\T & 
   \\
   y_1 &= \hv_1^\T (\uv + \vv) & 
   y_2 &= h_{21}\hv_1^\T \vv & 
   y_3 &= h_{31}\gv_1^\T \uv \\
   z_1 &= \gv_1^\T (\uv + \vv) &
   z_2 &= g_{21}\hv_1^\T \vv & 
   z_3 &= g_{31}\gv_1^\T \uv 
\end{flalign}
\end{subequations}
In the first slot, the transmitter sends the mixed signal to both users. In the second 
slot, the transmitter sends the interference seen by receiver 1 in the first slot. The role of this stage is two-fold: \emph{resolving
interference for user~1 and reinforcing signal for user~2}. In the third
slot, the transmitter sends the interference seen by user 2 to help the users the other way around. Therefore, 
this variant of the MAT alignment is composed of two phases:
i)~broadcasting the mixed signal, and ii)~multicasting the mutual
interference $(\hv_1^\T\vv, \gv_1^\T \uv)$. 
At the end of three slots, the observations at the receivers are given by
\begin{align}
  \begin{bmatrix} y_1\\y_2\\y_3 \end{bmatrix} &= \underbrace{\begin{bmatrix}
    \hv_1^\T \\ 0 \\ h_{31} \gv_1^\T 
  \end{bmatrix}}_{\text{rank}=2} \uv + \underbrace{\begin{bmatrix}
    \hv_1^\T \\ h_{21} \hv_1^\T \\ 0 
  \end{bmatrix}}_{\text{rank}=1} \vv, 
\shortintertext{and} 
 \begin{bmatrix} z_1\\z_2\\z_3 \end{bmatrix} &= \underbrace{\begin{bmatrix}
   \gv_1^\T \\ g_{21} \hv_1^\T \\ 0   
 \end{bmatrix}}_{\text{rank}=2} \vv + \underbrace{\begin{bmatrix}
   \gv_1^\T \\  0 \\ g_{31} \gv_1^\T  
 \end{bmatrix}}_{\text{rank}=1} \uv. 
\end{align}%
For each user, the useful signal lies in a
two-dimensional subspace while the interference is aligned in a
one-dimensional subspace. Since the latter is not completely included in
the signal subspace, it is readily shown that two degrees of freedom are
achievable in the three-dimensional time space, yielding $\frac{2}{3}$ as the
average degrees of freedom. This variant, although trivial from the original MAT scheme, 
is crucial to the integration of the current CSI, if there is any. 

\subsection{Integrating the Imperfect Current CSI}

Based on the above variant of the MAT scheme, we propose the following
two-stage scheme that integrates the estimates of the current CSI. 

\subsubsection*{Phase 1 - Precoding and broadcasting the mixed
signals}

As in the above MAT variant, we first mix the two signals
as $\xv_1 = \uv + \vv$, except that $\uv$ and $\vv$ are precoded
beforehand 
\begin{align}
  \uv &= \Wm \tilde{\uv},\quad \vv = \Qm \tilde{\vv} 
\end{align}%
where 
  $\Wm \defeq [\begin{matrix}\wv_1 & {\wv_2}\end{matrix}] \in
    \mathbb{C}^{m\times 2}$ and 
    $\Qm \defeq [\begin{matrix}\qv_1 & {\qv_2}\end{matrix}] \in
      \mathbb{C}^{m\times 2}$
are the precoding matrices; 
$\tilde{\uv}\triangleq [\tilde{u}_1 \quad \tilde{u}_2]^\T$ and
$\tilde{\vv}\triangleq [\tilde{v}_1 \quad \tilde{v}_2]^\T$ are input
signals of dimension $2$ for user~1 and user~2, respectively.
Furthermore, we suppose that $\tilde{\uv}$ and $\tilde{\vv}$ are mutually independent. 
In this paper, we restrict ourselves to orthogonal precoders, i.e.,
$\Wm^\H \Wm = \Id$ and $\Qm^\H \Qm = \Id$.  In particular, we align
$\wv_2$ and $\qv_2$ with the estimated channels $\hat{\gv}_1$ and
$\hat{\hv}_1$, respectively. That is, 
\begin{subequations}\label{NullSpace}
\begin{align}
\wv_1 \in \text{null}(\hat{\gv}_1), \quad {\wv_2} \in \text{span}(\hat{\gv}_1) \\
\qv_1 \in \text{null}( \hat{\hv}_1),  \quad  {\qv_2} \in \text{span}( \hat{\hv}_1) 
\end{align}
\end{subequations}
Let us define the covariance matrices $\Lambdam \defeq\E[\tilde{\uv} \tilde{\uv}^\H]$
and $\Phim \defeq\E[\tilde{\vv} \tilde{\vv}^\H]$. 
Without loss of generality, we can assume that  
both $\Lambdam$ and $\Phim$ are diagonal. Hence, the power constraint is
simply
\begin{align}
  \lambda_1+\lambda_2+\phi_1+\phi_2 &\le P.
\end{align}%
In other words, for each user, we send two streams in two
orthogonal directions: one aligned with the estimated channel while the
other one perpendicular to it.

\subsubsection*{{Phase 2 - Quantizing and multicasting the mutual interference
}} 

As the second phase of the MAT variant, the objective of this
phase is, by sending the mutual interferences $(\hv_1^\T\vv, \gv_1^\T
\uv)$ seen at the receivers, to resolve
the interference \emph{and} to reinforce the useful signal at the same
time. However, unlike the original MAT scheme where the interferences $(\hv_1^\T\vv,
\gv_1^\T \uv)$ is transmitted in an analog form, we will quantize it and then transmit the
digital version. The rationale behind this choice is as follows. With (imperfect)~CSI on the current
channel, the transmitter can use the precoding to align the signals and allocate the transmit
power in such a way that the mutual interferences have a reduced power,  
without sacrificing too much the received signal
power.\footnote{With no CSIT on the current channel, the only way to reduce
the interference power is to reduce the transmit power, therefore the
received signal power.} As a result, we should be able to save the
resource needed to multicast the interferences, which increases the
average rate. The reduction can be significant when the current CSI is
good enough. In this case, the analog transmission is not suitable any
more, due to the mismatch of the source power and available transmit
power. Therefore, a good alternative is to quantize the interferences before
transmission. The number of quantization bits depends naturally on the
interference power, which means that the multicasting can be done
efficiently.

Let us look into the interferences by taking into account the precoding. 
We start by examining the interference $\eta_1  \defeq \hv_1^\T \vv$
seen by user~1. It can be rewritten as
\begin{align}
  \eta_1 &= \hv_1^\T \Qm \tilde{\vv} \\
&= (\hv_1^\T \qv_1) \tilde{v}_1 + (\hv_1^\T {\qv_2}) \tilde{v}_2 \\
&= (\deltav_{1}^\T \qv_1) \tilde{v}_1 + (\hv_1^\T {\qv_2})
\tilde{v}_2 \label{eq:eta1}
\end{align}%
where $\deltav_{1}^\T \qv_1$ and $\hv_1^\T {\qv_2}$ 
are known at the end of the first slot to both
receivers, according to Assumption~\ref{assumption:CSI}. Therefore, the
average power of $\eta_1$ is $\sigma^2_{\eta_1} \defeq \E\bigl(
\abs{\eta_1}^2 \bigr)$, i.e., 
\begin{align}
  \sigma^2_{\eta_1} &=  \abs{\deltav_{1}^\T \qv_1}^2
  \phi_1 + \abs{\hv_{1}^\T {\qv}_2}^2
  \phi_2. \label{eq:ubip}
\end{align}%
Similarly, for the interference seen by user~2 during the first slot 
$\eta_2 \defeq \gv_1^\T \uv$, the average power is   
\begin{align}
  \sigma^2_{\eta_2} &=  \abs{\varepsilonv_{1}^\T \wv_1}^2
  \lambda_1 + \abs{\gv_{1}^\T \wv_2}^2
  \lambda_2. 
\end{align}%
Obviously, the interference powers $\sigma^2_{\eta_1}$ and
$\sigma^2_{\eta_2}$ depend on the both the precoder and the power
allocation at the transmitter. The power allocation issue will be
discussed in the next section.

The first step is to quantize $(\hv_1^\T\vv, \gv_1^\T \uv)$. 
Although it is possible to apply directly a 2-dimensional quantizer, we choose to quantize both
signals individually for simplicty of demonstration. Let us assume that
an $R_k$-bits scalar quantizer is used for $\eta_k$, $k=1,2$.
Hence, we have  
\begin{align}
  \eta_k &= \hat{\eta}_k + \xi_{\Delta,k}, \quad \hat{\eta}_k\in
  \mathcal{C}_k
\end{align}%
where $\mathcal{C}_k$, $k = 1,2$, is a quantization codebook of size $2^{R_k}$;
$\hat{\eta}_k$ and $\xi_{\Delta,k}$ are the quantized value and the
quantization noise, respectively. 
The indices of both $\hat{\eta}_1$ and $\hat{\eta}_2$,
represented in $R_1+R_2$ bits, are then multicast to both users in
$\kappa$ channel uses. As will be specified in the next section, we
choose $\kappa$ such
that the indices can be recovered with high probability.

At the receivers' side, each user first tries to recover $(\hat{\eta}_1, \hat{\eta}_2)$. 
If this step is done successfully, then receiver~1 has
\begin{align}
  y_1 &= \hv_1^\T \uv + \eta_1 + e_1 \\
  \hat{\eta}_1 &= \eta_1 - \xi_{\Delta,1} \\
  \hat{\eta}_2 &= \eta_2 - \xi_{\Delta,2} = \gv_{1}^\T \uv - \xi_{\Delta,2}
\end{align}%
from which an equivalent $2\times2$ MIMO channel is obtained
\begin{align}
	\tilde{\yv} \triangleq \begin{bmatrix} y_1 -\hat{\eta}_1
          \\\hat{\eta}_2 \end{bmatrix}  = { \Sm_1 } \Wm \tilde{\uv} +
            {\begin{bmatrix} e_1 + \xi_{\Delta,1} \\ -
              \xi_{\Delta,2} \end{bmatrix}} \label{eq:MIMO1}
\end{align}
where the noise $\tilde{\nv} \defeq [e_1 + \xi_{\Delta,1} \quad
-\xi_{\Delta,2}]^\T$ is not Gaussian and can depend on the signal
in general; the equivalent channel matrix is $\Fm\defeq { \Sm_1 }
\Wm \in \CC^{2\times2}$. Similarly, if receiver~2 can recover
$(\hat{\eta}_1, \hat{\eta}_2)$ correctly, then the following term is
available
\begin{align}
	\tilde{\zv} \triangleq \begin{bmatrix} \hat{\eta}_1 \\ z_1 - \hat{\eta}_2
        \end{bmatrix}  = { \Sm_1 } \Qm \tilde{\vv} +
            {\begin{bmatrix} - \xi_{\Delta,1} \\ b_1 + \xi_{\Delta,2}
            \end{bmatrix}}. 
              \label{eq:MIMO2}
\end{align}
In order to finally recover the message, each user performs the MIMO
decoding of the above equivalent channel.

\section{Achievable Degrees of Freedom}

\newcommand{\DoF}{\textsf{DoF}}

In this section, we analyze the achievable rate of the proposed scheme
in the high SNR regime. In particular, we are interested in the pre-log
factor of the achievable rate, the so-called degrees of freedom~(DoF).  
However, since we do not assume ergodic fading process in this work, 
we do not use directly ergodic capacity as our performance measure.
Instead, following the definition of multiplexing gain in \cite{Zheng_Tse}, 
we define the achievable degrees of freedom as follows. 
 
\begin{definition}[achievable degrees of freedom]
  For a family of codes $\{ \Xc(P) \}$ of length $L$ and rate
  $R(P)$ bits per channel use, we let $P_e(P)$ be the average probability
  of error and define
  \begin{align}
    r &\defeq \lim_{P\to\infty} \frac{R(P)}{\log P}. 
  \end{align}%
  Then, the achievable degrees of freedom of $\Xc$ is defined as 
  \begin{align}
    \DoF &\defeq \sup\left\{ r:\ \lim_{P\to\infty} P_e(P) =
    0\right\}. 
  \end{align}%
\end{definition}
\vspace{\baselineskip}
In other words, the DoF defined in this work is the maximum pre-log factor of
the rate of a coding scheme for a reliable communication in the high SNR
regime. Note that the code length $L$ here is fixed, which avoids the
involvement of the whole fading process.

In the following, we focus on the symmetrical case where the two users have
the same data rate. The whole achievable region is straightforward following
the same lines. In addition, we assume that $\displaystyle
\lim_{P\to\infty} \alpha_P$ exists and define
\begin{align}
\alpha \defeq \lim_{P\to\infty} \alpha_P.  
\end{align}%
The main result is stated in the following theorem. 
\begin{theorem} \label{theorem:betterDoF}
  In the two-user MISO broadcast channel with delayed perfect CSIT and
  imperfect current CSIT~(Assumption \ref{assumption:CSI}), 
  the following DoF is achievable for each user
\begin{align}
  d = \begin{dcases}
\frac{2-\alpha}{3-2\alpha},& \alpha \in[0,1] \\
1, & \alpha > 1.
  \end{dcases}
\label{eq:d1d2}
\end{align} 
\end{theorem}
\vspace{0.5\baselineskip}

\begin{figure}
 \centering
\includegraphics[width=0.95\columnwidth]{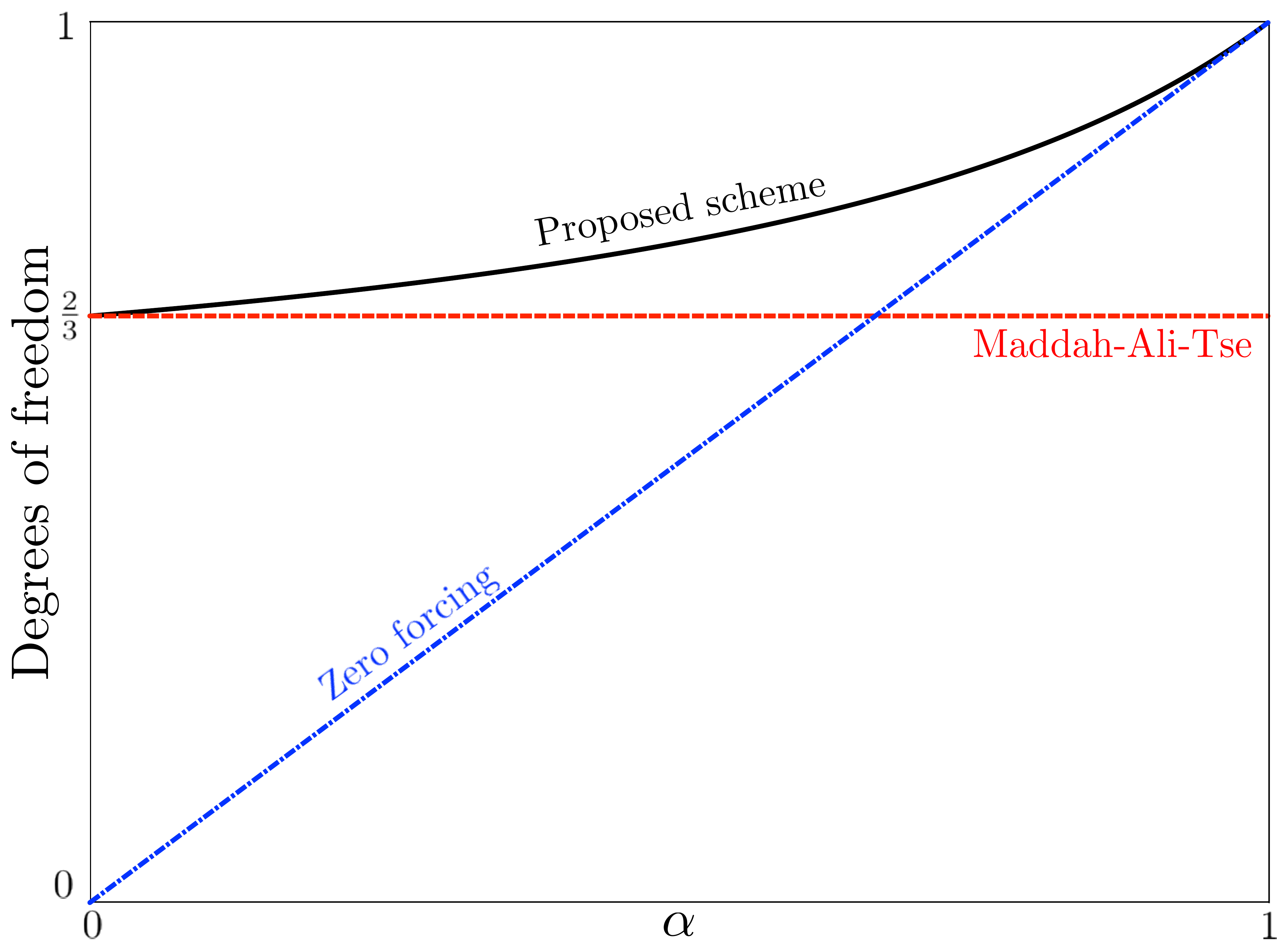}
\caption{Comparison of the achievable DoF between the proposed scheme and the zero-forcing and MAT alignment as a function of $\alpha$.}
\label{fig:DoF}
\end{figure}

Note that when $\alpha$ is close to $0$, the estimation of current CSIT is bad 
and therefore useless. In this case, the optimal scheme is MAT~\cite{maddah2010degrees}, achieving DoF of $\frac{2}{3}$ for each user. 
On the other hand, when $\alpha\ge1$, the estimation is good and the interference 
at the receivers due to the imperfect
estimation is below the noise level and thus can be neglected as far as
the DoF is concerned. In this case, ZF with the estimated 
current CSI is asymptotically optimal, achieving degrees of freedom $1$ for each user. 
Interestingly, our result~(Fig.~\eqref{fig:DoF}) reveals that strictly larger DoF than
$\max\{\frac{2}{3}, \alpha\}$ can be
obtained by exploiting both the imperfect current CSIT and the perfect delayed CSIT 
in an intermediate regime $\alpha\in(0,1)$. The intuition behind equation~\eqref{eq:d1d2} is as follows. 
Decreasing the interference power will reduce the receive power of
useful signal, incurring a loss of degrees of freedom. On the other
hand, decreasing the interference power will also save the resources needed
to communicate the interference \emph{a la} MAT. By smartly
aligning the signals and allocating the transmit power, the proposed
scheme loses only $\alpha$~(numerator in \eqref{eq:d1d2}) degrees of
freedom, but reduces $2\alpha$ channel uses~(denominator in
\eqref{eq:d1d2}). 

The rest of the section is devoted to the proof of the Theorem.
Some important ingredients of the proposed scheme are:
\begin{itemize}
  \item Two independent Gaussian codebooks $\Xc_1$ and $\Xc_2$ with same
    size $2^R$ are used for $\tilde{\uv}$ and $\tilde{\vv}$, respectively.
  \item Since we are interested in the symmetrical case, same power
    allocation scheme is applied to both user, i.e.,
    $\phi_l = \lambda_l = P_l$, $l=1,2$. Hence, we have $P_1+P_2 = P/2$.
  \item Truncated uniform quantization with unit step and
    truncation value $\displaystyle\bar{\eta} = P^{\frac{1+\zeta}{2}}
    \sigma $, for some $\zeta>0$, is used for both the real and
    imaginary parts of $\eta_1$ and $\eta_2$, i.e.,
    \begin{align}
      \hat{\eta}_k &= \left[ \text{trunc} (\Re[\eta_k]) \right] +
      i\,\left[ \text{trunc} (\Im[\eta_k]) \right]
    \end{align}%
    where $\text{trunc}(x)=x$ if $x\in[-\bar{\eta},
    \bar{\eta}]$ and $0$ otherwise. 
  \item The double indices of $(\hat{\eta}_1, \hat{\eta}_2)$,
    represented in 
    \begin{align}
      4 \log(2\lceil\bar{\eta}\rceil) &\approx 4 + 2(1+\zeta-\alpha_P)
      \log P \quad \text{bits},
    \end{align}%
    are sent with a multicast code.  
\end{itemize}

We define the error event $\Ec$ as the event that one of the users
cannot recover his message correctly. 
It can be shown that this event implies one of the following events:  
\begin{itemize}
  \item Quantization range error $\Ec_{\Delta}$: the amplitude of
    real or imaginary parts of interferences is out of $[-\bar{\eta},
    \bar{\eta}]$;
  \item Multicast error $\Ec_{\textsf{mc}}$: one of the users
    cannot recover the double indices of $(\hat{\eta}_1, \hat{\eta}_2)$
    correctly;
  \item MIMO decoding error $\Ec_{\textsf{mimo}}$: based on the received
    signal and the recovered indices, one of the users cannot recover
    his original message after performing a MIMO decoding of the equivalent
    channel~\eqref{eq:MIMO1} or~\eqref{eq:MIMO2}.
\end{itemize}
That is, $\Ec \subseteq \Ec_{\Delta} \cup \Ec_{\textsf{mc}} \cup
\Ec_{\textsf{mimo}}$.
Therefore, we have 
\begin{align}
  P_e \le \P[\Ec_{\Delta}] + \P[\Ec_{\textsf{mc}}] + \P[
  \Ec_{\textsf{mimo}} \cap \bar{\Ec}_{\Delta} \cap \bar{\Ec}_{\textsf{mc}} ].
\end{align}%
In the following, we examine the individual error events.

\subsubsection{Quantization range error $\Ec_{\Delta}$}

This event is the union of the four events: 
$\{\Re[ \eta_1 ]  > \bar{\eta} \}$, $\{ \Im[ \eta_1 ] >
\bar{\eta} \}$, $\{ \Re[ \eta_2 ] > \bar{\eta} \}$, and $\{ \Im[ \eta_2 ]
> \bar{\eta} \}$. This event implies that the quantization error is not bounded. 
From \eqref{eq:eta1} and the fact that Gaussian
codebooks are used, $\eta_k \sim \CN[0,\sigma_{\eta_k}^2]$, i.e.,
$\Re[ \eta_k ], \Im[ \eta_k ] \sim \Nc(0,\frac{\sigma_{\eta_k}^2}{2})$,
$k=1,2$, conditional on the channel states. We can show that~(cf. Appendix), for any $\epsilon>0$,
\begin{multline}
   {\P[\Re[ \eta_1 ] > \bar{\eta}] } \le e^{-P^{\epsilon}} +
  \frac{1}{4m^2} P^{-2(\zeta-\epsilon)} \\   + \frac{1}{4m^2} P^{-2(\zeta-\epsilon+1-\alpha_P-\beta_P)}
  \label{eq:tmp876}
\end{multline}%
where we define 
\begin{align}
  \beta_P &\defeq \frac{\log P_2}{\log P}. 
\end{align}%
Note that, due to the symmetry, the probabilities for the four
events, i.e., $\P[\Re[ \eta_1 ]  > \bar{\eta}]$,
${\P[\Im[ \eta_1 ] >
\bar{\eta}] }$, ${\P[\Re[ \eta_2 ] > \bar{\eta}] }$, and 
${\P[\Im[ \eta_2 ] > \bar{\eta}] }$, have the same upper bound
\eqref{eq:tmp876}. Therefore, by the union bound, we have $\P[\Ec_\Delta]\le 4\, \P[\Re[ \eta_1 ] > \bar{\eta}]$. 
From \eqref{eq:tmp876}, a sufficient condition for $\displaystyle \lim_{P\to\infty}
\P[\Ec_\Delta] = 0$ is $\zeta>\epsilon>0$ and 
$\displaystyle \lim_{P\to\infty} 1-\alpha_P - \beta_P \ge 0$, i.e.,
\begin{align}
  \lim_{P\to\infty} \beta_P \le 1-\alpha, \label{eq:beta}
\end{align}%
meaning that the power $P_2$ should not scale faster than
$P^{1-\alpha}$.

\subsubsection{Multicast error $\Ec_{\textsf{mc}}$}

First, we provide the following lemma proved in the Appendix. 
\begin{lemma}\label{lemma:DoF-MC}
The DoF of the multicast communication in the considered two-user MISO
channel is $\DoF_{\textsf{mc}} = 1$. That is, for any $\delta>0$, there
exists a code with rate $(1-\delta) \log P$, such that
the average error probability goes to $0$ when $P\to\infty$.
\end{lemma}

Note that the number of bits needed to describe the indices is $4+2(1+\zeta-\alpha_P) \log P$. From
Lemma~\ref{lemma:DoF-MC}, we know that for any $\delta<0$,
a rate $(1-\delta)\log P$ can be achieved reliably
when $P\to\infty$. Therefore, as long as the number of channel uses  
\begin{align}
  \kappa  &\ge \frac{4}{(1-\delta)\log P} +
  \frac{2(1+\zeta-\alpha_P)}{1-\delta},  \label{eq:L}
\end{align}%
we can guarantee that $P\left\{ \Ec_1 \right\} \to 0$ when $P\to\infty$.

\subsubsection{MIMO decoding error $\Ec_{\textsf{mimo}}$}
Let $\Ec_{{\textsf{mimo},k}}$ be the MIMO decoding error at
receiver~$k$, $k=1,2$. It is obvious that $\P[\Ec_{\textsf{mimo}}] \le
\P[\Ec_{\textsf{mimo},1}] + \P[\Ec_{\textsf{mimo},2}]$. Due to the
symmetry, we can focus on $\Ec_{\textsf{mimo},1}$. First, we introduce $\epsilon'>0$ and define 
\begin{align}
\Oc_{\epsilon'} &\defeq \left\{ \Fm:\ \log\det[\Id + \Fm \Lambdam
\Fm^\H] < R + \epsilon' \log P\right\}. \label{eq:Oc}
\end{align}%
Therefore, the error probability can be upper-bounded by 
\begin{align}
  \MoveEqLeft[3] {\P[\Ec_{\textsf{mimo},1} \cap \bar{\Ec}_{\Delta} \cap \bar{\Ec}_{\textsf{mc}}]}\nonumber \\ 
  &\le {\P[\Ec_{\textsf{mimo},1} \cap \bar{\Ec}_{\Delta} \cap
  \bar{\Ec}_{\textsf{mc}} \cap \bar{\Oc}_{\epsilon'}]} +
  \P[\Oc_{\epsilon'}].
\end{align}%
It can be shown~(cf. Appendix) that 
\begin{align}
\P[\Ec_{\textsf{mimo},1} \cap \bar{\Ec}_{\Delta} \cap
\bar{\Ec}_{\textsf{mc}} \cap \bar{\Oc}_{\epsilon'}] &\leq 32 P^{-\epsilon'} \label{eq:Pe}
\end{align}%
and that $\displaystyle \lim_{P\to\infty} \P[\Oc_{\epsilon'}] = 0$ for any 
\begin{align}
  r &\le 1 + \beta_P - \epsilon' - \epsilon'' \label{eq:r}
\end{align}%
and $\epsilon''>0$. 

From \eqref{eq:L} and \eqref{eq:r}, the proposed
scheme can deliver \emph{reliably} 
\begin{align}
  \frac{ r }{1 + \kappa } \log P 
  &\le \frac{ (1-\delta)(1+\beta_P-\epsilon'-\epsilon'') }{1 - \delta + \frac{4}{\log
   P} + 2(1+\zeta-\alpha_P)} \log P
\end{align}%
bits per channel use, when $P\to\infty$,
from which we can deduce the achievable pre-log factor 
\begin{align}
  \frac{\displaystyle (1-\delta)(1+ \lim_{P\to\infty}\beta_P -\epsilon'-\epsilon'')}{1 - \delta
  + 2(1+\zeta-\alpha)}. \label{eq:tmp343}
\end{align}%
We can maximize \eqref{eq:tmp343} over the power exponent
$\beta_P$ under the constraint \eqref{eq:beta}. The maximizing value of
$\displaystyle \lim_{P\to\infty}\beta_P$ is $1-\alpha$,
i.e., the power attributed to the stream in the direction of estimated
channel should scale as $P^{1-\alpha}$.
Finally, by making $\zeta$, $\epsilon'$, $\epsilon''$, and
$\delta$ as close to $0$ as possible in~\eqref{eq:tmp343}, we prove the achievable DoF
for user~1, given in \eqref{eq:d1d2}. Due to the symmetry, same
proof applies to finding precisely the same DoF for user~2. \QED

\section{Example: Doppler Fading Process}

The main result on the achievable DoF has been presented in terms of an artificial parameter $\alpha$, denoting the speed of decay of the estimation error $\sigma^2\sim P^{-\alpha}$ in the current CSIT. In this section, we provide an example showing the practical interpretation
of this parameter. Focusing on receiver 1 due to symmetry, we describe the fading process, channel estimation, and feedback
scheme as follows: 
\begin{itemize}
  \item The channel fading $\hv_t$ follows a Doppler
    process with power spectral density $S_h(w)$. The channel coefficients
    are strictly band-limited to $[-F,F]$ with $F=\frac{v f_c 
    T_f}{c} < \frac{1}{2}$ where $v, f_c, T_f, c$ denotes the mobile speed
    in m/h, the carrier frequency in Hz, the slot duration in sec, the light
    speed in m/sec.
  \item The channel estimation is done at the receivers side with
    pilot-based downlink training. At slot $t$, receiver 1
    estimates $\hv_{t}$ based on a sequence of the noisy observations $\{\sv_{\tau} = \sqrt{P\gamma} \hv_{\tau} + \nuv_{\tau}\}$ up to $t$, where a constant $\gamma\geq 1$ denotes the resource factor dedicated to the training and $\nuv_t$ is AWGN with zero mean unit covariance. The
    estimate is denoted by $\tilde{\hv}_t$ with  
    \begin{align}
      \hv_t &= \tilde{\hv}_t + \tilde{\deltav}_t \label{eq:estimation}
    \end{align}%
    Under this model, the estimation error vanishes as
    $\E[\norm{\tilde{\deltav}_t}^2] \sim P^{-1}$.
  \item At the end of slot $t$, the noisy observation 
    $\sv_t$ is sent to the transmitter and receiver 2 over a noise-free channel. 
    At slot $t+1$, based on the noisy observation $\{\sv_{\tau}\}$ up to $t$, 
     the transmitter and receiver 2 acquire the prediction ${\hat{\hv}}_{t+1}$ of $\hv_{t+1}$ and estimation $\tilde{\hv}_t$ of $\hv_{t}$. The corresponding prediction model is  
    \begin{align}
      {\hv}_{t} &= {\hat{\hv}}_{t} +
      {\hat{\deltav}}_{t}\label{eq:prediction}
    \end{align}%
    From \cite[Lemma 1]{caire2010multiuser}, we have 
    $\E[\norm{{\hat{\deltav}}_t}^2] \sim P^{-(1-2F)}$. 
\end{itemize}

In this channel with imperfect delayed CSIT, we can still apply the
proposed scheme and analysis in exactly the same way as above except for 
the following principal changes. First, the
known interference becomes $\eta_1 = \tilde{\hv}_1^\T \vv$ and $\eta_2 =
\tilde{\gv}_1^\T \uv$ and the received signal $y_1$ becomes $y_1 =
\tilde{\hv}_1^\T\uv + \eta_1 + e_1 + \tilde{\deltav}_1^\T \xv_1$. Second,
the precoding is now based the prediction, still given by
\eqref{NullSpace}. Last,
the parameter $\alpha$, charactering the estimation error 
$\deltav_1$ of the current channel states in Assumption~\ref{assumption:CSI}, now
characterizes the mismatch between the estimated CSIT and the predicted
one. That is, $\E[\norm{\tilde{\deltav_1} - {\hat{\deltav}}_1}^2] \sim
P^{-\alpha}$, which means that, from \eqref{eq:estimation} and
\eqref{eq:prediction}, 
$\alpha = 1-2F$. 
Consequently, the equivalent MIMO channel~\eqref{eq:MIMO1} becomes
\begin{align}
  \tilde{\yv}  = { \tilde{\Sm}_1 } \Wm \tilde{\uv} +
            {\begin{bmatrix} e_1 + \xi_{\Delta,1} + \tilde{\deltav}_1^\T \xv_1 \\ -
              \xi_{\Delta,2} \end{bmatrix}}. 
\end{align}
 Since it can be shown that
$\P[\abs{\tilde{\deltav}_1^\T \xv_1}^2>P^{\epsilon}] < O(P^{-\epsilon})$
from the Chebyshev's inequality, $\tilde{\deltav}_1^\T \xv_1$ can be
considered bounded as far as the DoF is concerned and thus does not
affect the achievable DoF. 

\section{Conclusion}
We considered a practical scenario of the time-correlated MISO broadcast
channel where the transmitter takes an opportunity to exploit both past
(delayed) channel state and an estimate of current channel state. We
proposed a novel multi-slotted strategy which enhances the degrees of
freedom promised by the MAT scheme according to the quality of the
current channel knowlege. The optimality of this scheme in terms of
degrees of freedom remains unknown. Inner and outer bounds of the DoF
region as well as extensions to more general cases and other network
models is under investigation and will be reported in the full paper
\cite{FullVersion}.

\appendix
\subsection{Proof of Equation \eqref{eq:tmp876}}
First, we have
\begin{align}
  \P[\Re[ \eta_1 ] > \bar{\eta}] &= \E[Q\left(
  \frac{\bar{\eta}_1}{{\sigma_{\eta_1} / \sqrt{2}}} \right)]  \\
  &\le \E[\exp\left( - \frac{\bar{\eta}_1^2}{\sigma_{\eta_1}^2}\right)]
  \label{eq:tmp892}\\
  &\le \E[\exp\left( - \frac{\bar{\eta}_1^2}{A}\right)]
\end{align}%
where the first equality comes from the Gaussian distribution
conditional on the channel states; to obtain \eqref{eq:tmp892}, we
applied $Q(x) \le e^{-x^2/2}$; the last inequality is from the fact that
$\sigma_{\eta_1}^2 \le A$ with $A\defeq { \norm{\deltav_{1}}^2
P + \norm{{\hv}_1}^2 P_2 }$. Now, we can go further with the upper bound, by introducing
$\epsilon>0$, 
\begin{align}
  \lefteqn{\P[\Re[ \eta_1 ] > \bar{\eta}] } \nonumber \\ 
  &\le \P[ A \le \bar{\eta}^2 P^{-\epsilon} ] e^{-P^{\epsilon}}  + \P[ A > \bar{\eta}^2 P^{-\epsilon} ]  \\
  &\le e^{-P^{\epsilon}} + \P[ A > \bar{\eta}^2 P^{-\epsilon} ] \\
  &\le e^{-P^{\epsilon}} + \P[ \norm{\deltav_{1}}^2 P >
  \frac{1}{2} \bar{\eta}^2 P^{-\epsilon} ] \nonumber \\ 
  &\qquad + \P[ \norm{{\hv}_1}^2 P_2 > \frac{1}{2} \bar{\eta}^2
  P^{-\epsilon} ]  \label{eq:union_bound} \\
  &= e^{-P^{\epsilon}} + \P[
  \frac{\norm{\deltav_{1}}^2}{m\sigma^2} >
  \frac{1}{2m} P^{\zeta-\epsilon} ] \nonumber \\ 
  &\qquad + \P[ \frac{\norm{{\hv}_1}^2}{m} > \frac{1}{2m} 
  P^{\zeta-\epsilon+1-\alpha_P-\beta} ] \\
  &\le e^{-P^{\epsilon}} + \frac{1}{4m^2} P^{-2(\zeta-\epsilon)} + \frac{1}{4m^2} P^{-2(\zeta-\epsilon+1-\alpha_P-\beta_P)}
\end{align}%
where \eqref{eq:union_bound} is from the union bound;  the last inequality is Chebyshev's inequality. 

\subsection{Proof Outline of Lemma~\ref{lemma:DoF-MC}}
  Since each user has only one antenna, the DoF per user for the
  multicast communication is upper-bounded by $1$. For the lower bound,
  let us consider a trivial scheme in which only one transmit antenna
  out of $m$ is used. The MISO BC becomes a SISO BC that is degraded.
  And the multicast capacity is just that of the worse user, which
  yields $1$ as DoF as well. This can be achieved with a single-letter
  code~(e.g., QAM constellation). 
  \QED

\subsection{Proof of Equation \eqref{eq:Pe}}

By applying the union bound, we have
\begin{align}
\MoveEqLeft \P\bigl(\Ec_{\textsf{mimo},1} \cap
\overbrace{\bar{\Ec}_{\Delta} \cap \bar{\Ec}_{\textsf{mc}} \cap
\bar{\Oc}_{\epsilon'}}^{\Bc} \bigr)  \nonumber \\
&\leq
\P\bigl( (W_1 \ne \hat{W}_1) \cap {\Bc} \bigr) +
\P\bigl( (W_2 \ne \hat{W}_2) \cap {\Bc} \bigr) 
\end{align}%
where $W_k$ is the original message for user~$k$, while $\hat{W}_k$ is
the decoded message based on the received and reconstructed
observations, $k=1,2$. For simplicity, we assume that minimum Euclidean distance decoding is
used\footnote{Since the noise is not Gaussian and depends on the signal
in general, it does not correspond to maximum likelihood decoding.}.  
Let us focus on the error event for user~1, i.e., 
$\P[(W_1 \ne \hat{W}_1) \cap {\Bc}]$. To that end, we look into the
pair-wise error probability for a pair of different codewords
$\tilde{\uv}(0), \tilde{\uv}(1) \in \Xc_1$, denoted by 
$\P[\tilde{\uv}(0) \to \tilde{\uv}(1)]$. For a given channel realization
$\Fm$, we have
\begin{align}
  \MoveEqLeft {\P[(\tilde{\uv}(0) \to \tilde{\uv}(1)) \cap {\Bc} \cond \Fm ]} \nonumber \\
  &\le \P[ \left( \frac{\norm*{\Fm (\tilde{\uv}(0)-\tilde{\uv}(1))}}{2}
  \le \norm{\tilde{n}} \right) \cap {\Bc} ] \\
  &\le \P[ \left( \norm*{{\Fm \tilde{\uv}_{d}}}^2 \le 2 (
  \abs{e_1}^2 + \abs{\xi_{\Delta,1}}^2 + \abs{\xi_{\Delta,2}}^2 )\right)
  \cap {\Bc}    ] \nonumber \\
  &\le \P[ \left(\norm*{{\Fm \tilde{\uv}_{d}}}^2 \le 2 \abs{e_1}^2 +
  1\right) \cap {\Bc} ] \label{eq:tmp812} \\
  &\le \P[ \left(\norm*{{\Fm \tilde{\uv}_{d}}}^2 \le 2 \abs{e_1}^2 + 1\right) \cap {\bar{\Oc}_{\epsilon'}} ] \\
  &\le \P[ \norm*{{\Fm \tilde{\uv}_{d}}}^2 \le 2 \abs{e_1}^2 + 1 ] \one_{\bar{\Oc}_{\epsilon'}}\!(\Fm) 
\end{align}%
where $\one_{\bar{\Oc}_{\epsilon'}}\!(\Fm)$ is the indicator function
that gives $1$ if $\Fm \in\bar{\Oc}_{\epsilon'}$ and $0$ otherwise;
\eqref{eq:tmp812} is from the fact that the quantization error
$\abs{\xi_{\Delta,k}}$ is bounded by $\frac{1}{2}$, $k=1,2$.
We can go further with the probability term
\begin{align}
  \MoveEqLeft \P[ \left(\norm*{{\Fm \tilde{\uv}_{d}}}^2 \le 2 \abs{e_1}^2 +
  1\right) ] \\
  &\le \P[ \norm*{{\Fm \tilde{\uv}_{d}}}^2 \le 4 \abs{e_1}^2 ] + \P[
  \norm*{{\Fm \tilde{\uv}_{d}}}^2 \le 2 ] \\
  &\le \E[ \exp\left( -\frac{1}{4} \norm{ \Fm \tilde{\uv}_{d}}^2
  \right) ] + \P[ \norm*{{\Fm \tilde{\uv}_{d}}}^2 \le 2 ] \\
  &\le \det[\Id + \frac{1}{4} \Fm \Lambdam \Fm^\H]^{-1} \nonumber \\
  &\qquad + \P[ \rho_1 \le \frac{2}{\mu_1}] \P[ \rho_2 \le \frac{2}{\mu_2}]  \\
  &\le 16 \, {\det[\Id + \Fm \Lambdam \Fm^\H]}^{-1} + (1-e^{-\frac{2}{\mu_1}} )( 1-e^{-\frac{2}{\mu_2}} )  \\
  &\le 16 \, {\det[\Id + \Fm \Lambdam \Fm^\H]}^{-1}  +
  \frac{16}{(2+\mu_1)(2+\mu_2)}  \label{eq:ubexp}\\ 
  &= 32\, {\det[\Id + \Fm \Lambdam \Fm^\H]}^{-1}
\end{align}%
where $\tilde{\uv}_d\defeq (\tilde{\uv}_0 -
\tilde{\uv}_1)/\sqrt{2} \sim \CN(0,\Lambdam)$; $\norm*{{\Fm
\tilde{\uv}_{d}}}^2 \overset{d}{=} \mu_1 \rho_1 + \mu_2 \rho_2$ with
$\rho_1, \rho_2\sim\exp(1)$ and $\mu_1\ge\mu_2$
being the two eigenvalues of $\Fm \Lambdam \Fm^\H$; \eqref{eq:ubexp} is
obtained by applying 
\begin{align}
  1-\exp\Bigl(-\frac{1}{x}\Bigr)\le \frac{2}{1+x}, \quad \text{for }
  x\ge 0. 
\end{align}%
Applying the union bound on all possible codewords pairs, we finally
obtain
\begin{align}
  \MoveEqLeft \P[(W_1 \ne \hat{W}_1) \cap {\Bc}] \nonumber \\
  &\le 2^R \, \E_{\Fm}\biggl( {\P[(\uv(0) \to \uv(1)) \cap {\Bc} \cond \Fm ]}\biggr) \\ 
  &\le 32 \, P^r \, \E_{\Fm}\biggl( {\det[\Id + \Fm \Lambdam \Fm^\H]}^{-1} \, \one_{\bar{\Oc}_{\epsilon'}}\!(\Fm) \biggr) \\ 
  &\le 32 \, P^r \,  \E_{\Fm}\biggl( P^{-(r+\epsilon')} \biggr) \\ 
  &= 32 \, P^{-\epsilon'} 
\end{align}%
where we used the fact that $\det[\Id + \Fm \Lambdam \Fm^\H]^{-1} \le
P^{-(r+\epsilon')}$ for any $\Fm \in \bar{\Oc}_{\epsilon'}$, from the
definition~\eqref{eq:Oc}.

\subsection{Proof of Equation \eqref{eq:r}}

The probability $\P[\Oc_{\epsilon'}]$ is upper-bounded as follows
\begin{align}
  \P[\Oc_{\epsilon'}] &\le \P[ \det[\Fm \Lambdam \Fm^\H] <
  P^{r+\epsilon'} ] \\
  &= \P[ \det[\Fm \Fm^\H] \det[\Lambdam]  < P^{r+\epsilon'} ] \\
  &= \P[ \det[\Fm \Fm^\H] < P^{r-1-\beta_P+\epsilon'} ].
\end{align}%
We know that for any $m\times 2$ matrix $\Am = [\av_1 \ \av_2]$, we
have 
\begin{align}
 \det[\Am^\H \Am] = \norm{\av_2}^2 \av_1^\H (\Id - \bar{\av}_2 \bar{\av}_2^\H) \av_1
\end{align}%
with $\bar{\av} = \frac{\av }{\norm{\av}}$. We identify $\Fm$ with $\Am^\H$ and obtain
$\norm{ \av_2}^2 = \norm{\gv_1^\T \Wm }^2 = \abs{\gv_1^\T \wv_1}^2
+ \abs{\gv_1^\T \wv_2}^2  \ge \abs{\gv_1^\T \wv_2}^2$ and $\av_1^\H = \hv_1^\T \Wm$
giving
\begin{align}
{\av}_1^\H (\Id - \bar{\av}_2\bar{\av}_2^\H)  {\av}_1  &= \hv_1^\T \Wm
(\Id - \bar{\av}_2\bar{\av}_2^\H) \Wm^\H \hv_1^*  \\
&= \hv_1^\T {\Mm} \hv_1^*  \\
&= \abs{ \hv_1^\T \wv_{\max} }^2 
\end{align}
where $\Mm \defeq \Wm (\Id - \bar{\av}_2\bar{\av}_2^\H) \Wm^\H$ that can
be readily shown to be rank one with a single non-zero eigenvalue $1$
corresponding to the eigen-vector $\wv_{\max}$; the last equality comes
from the fact that $\hv_1$ and $\wv_{\max}$ are independent. Therefore,
we have, by defining $\epsilon''\defeq - (r-1-\beta_P+\epsilon')$,
\begin{align}
  \MoveEqLeft[0] \P[ \det[\Fm \Fm^\H] < P^{-\epsilon''} ] \nonumber \\
  &\le \P[ \norm{\av_2}^2 \av_1^\H (\Id - \bar{\av}_2 \bar{\av}_2^\H) \av_1 < P^{-\epsilon''} ] \\
  &\le \P[ \abs{\gv_1^\T \wv_2}^2  \abs{ \hv_1^\T \wv_{\max} }^2 <
  P^{-{\epsilon''}} ] \\
  &\le \P[ \abs{\gv_1^\T \wv_2}^2  < P^{-\frac{\epsilon''}{2}} ] + \P[ \abs{ \hv_1^\T \wv_{\max} }^2 < P^{-\frac{\epsilon''}{2}} ].
   \label{eq:tmp329} 
\end{align}%
It can be shown that, as long as $\epsilon''>0$, both probabilities in
\eqref{eq:tmp329} goes to $0$ when $P\to\infty$.

\end{document}